\documentclass[12pt,preprint]{aastex}

\begin{document}

\title{Detecting the stochastic gravitational wave background using pulsar timing}

\author{Fredrick A. Jenet\altaffilmark{1}, George B. Hobbs\altaffilmark{2}, K.J. Lee\altaffilmark{3}, Richard N. Manchester\altaffilmark{2}}
\altaffiltext{1}{Center for Gravitational Wave Astronomy, University of Texas at Brownsville, TX 78520 (merlyn@alum.mit.edu)}
\altaffiltext{2}{Australia Telescope National Facility, CSIRO,  P.O.~Box~76, Epping, NSW~1710, Australia}
\altaffiltext{3}{Department of Astronomy, Peking University, Beijing 100871, China}

\begin{abstract}

The direct detection of gravitational waves is a major goal of current
astrophysics. We provide details of a new method for detecting a
stochastic background of gravitational waves using pulsar timing
data. Our results show that regular timing observations of 40 pulsars
each with a timing accuracy of 100\,ns will be able to make a direct
detection of the predicted stochastic background from coalescing black
holes within five years. With an improved pre-whitening algorithm, or
if the background is at the upper end of the predicted range, a
significant detection should be possible with only 20 pulsars.
\end{abstract}


\section{Introduction}

Analysis of pulsar pulse time-of-arrival (TOA) data shows that
pulsars, especially millisecond pulsars (MSPs), are very stable
clocks. Measurement of timing residuals, that is, the differences
between observed and predicted TOAs, enables the direct detection of
gravitational waves (GWs) \citep{ew75,saz78,det79}. The fluctuating
TOAs induced by a GW will be correlated between widely-spaced
pulsars. \citet{hd83} attempted to detect this correlation by
cross-correlating the time derivative of the timing residuals for
multiple pulsars.  In our work, we have developed a similar
cross-correlation technique and have, for the first time, a fully
analyzed method for combining multiple pulsar observations in order to
make an unambiguous detection of a GW background. We emphasize that,
in contrast to \citet{hd83}, our method is based entirely on the
measured residuals.

Only the effects of a stochastic background of GWs are considered.
Astrophysical sources of such a background include cosmological
processes \citep[e.g.][]{mag00} and coalescing massive black hole
binary systems \citep{jb03,wl03,ein+04}. We show that a direct
detection of a stochastic GW background is possible using pulsar
timing observations and that the significance of the detection depends
upon the number of pulsars observed, the root-mean-square (RMS) timing
noise achieved, the number of observations, and the power spectrum of
the measured timing residuals. The results are applied to the case of
the Parkes pulsar timing array
(PPTA\footnote{http://www.atnf.csiro.au/research/pulsar/psrtime}).

In the next section, the analysis technique is described. In \S 3
the significance of detecting a given stochastic background using this
method is estimated. The effects of pre-filtering the residual time
series are also discussed. The results are summarized in
\S 4.

\section{Detection Technique}
As a first step, the power spectra of the pulsar timing residuals are
analyzed.  If they all show a very red power-law spectrum, the
residuals may be dominated by a GW background. However, such red
spectra can also be due to period noise intrinsic to the pulsar,
uncorrected interstellar delays, inaccuracies in the Solar-System
ephemeris, or variations in terrestrial time standards
\citep[e.g.][]{fb90}. A GW background produces a unique signature in
the timing residuals which can only be confirmed by observing
correlated signals between multiple pulsars widely distributed on the
sky.

The presence of a stochastic GW background will cause the pulse TOAs
to fluctuate randomly, but these fluctuations will be correlated between
different pulsars. In order to detect the presence of a GW background,
one needs to first calculate the correlation coefficient
between the observed timing residuals of each pair of observed
pulsars:
\begin{equation}
r(\theta) = \frac{1}{N}\sum_{i=0}^{N-1} R(t_i,\hat{k}_1) R(t_i,\hat{k}_2) \label{r_stat}
\end{equation}
where $R(t_i,\hat{k})$ is the time series of $N$ pulsar residuals
sampled regularly in time, $\hat{k}_1$ and $\hat{k}_2$ are the
directions to the two pulsars, and $\cos(\theta) = \hat{k}_1 \cdot
\hat{k}_2$. It will be assumed that $R$ has zero mean and that each
pulsar pair has a unique angular separation.  $r(\theta)$ is written
only as a function of the angular separation since the GW background
is expected to be isotropic. In the presence of an isotropic GW
background, the ensemble-averaged value of $r(\theta)$ is given
by\footnote{For an outline of the calculation of $\zeta$ see
\citet{hd83}.}:
\begin{eqnarray}
\langle r(\theta)\rangle &=& \sigma_g^2 \zeta(\theta)  \\
\zeta(\theta)   &=& \frac{3}{2} x\log(x) -\frac{x}{4} + \frac{1}{2} +
\frac{1}{2}\delta(x) 
\end{eqnarray}
where $x = (1 - \cos(\theta))/2$, $\sigma_g$ is the RMS of the timing residuals
induced by the stochastic GW background, and $\delta(x)$ equals 1 for
$x=0$ and 0 otherwise. The detection technique proposed here simply
looks for the presence of the function $\zeta(\theta)$ in the measured
correlation coefficients $r(\theta)$.

Since one cannot perform the ensemble average in practice, the measured
statistic, $r(\theta)$, will be of the form $r(\theta) = \langle r(\theta)\rangle  + \Delta r(\theta)$,
where $\Delta r(\theta)$ is a ``noise term''. Since $r(\theta)$ is
calculated by summing over a large ($\geq 20$) number of data points,
$\Delta r(\theta)$ will be a Gaussian random variable for practical
purposes. The optimal way to detect the presence of a known functional
form within random data is to calculate the correlation between the
data and the known function. Hence, to detect the presence of
the GW background one needs to calculate
\begin{equation}
\rho = \frac{\frac{1}{N_p}\sum_{i=0}^{N_p-1} (r(\theta_i)-\bar{r})(\zeta(\theta_i)-\bar{\zeta})}{\sigma_r \sigma_\zeta} \label{rho_def}
\end{equation}
where $\theta_i$ is the angle between the $i$th pair of pulsars and
$N_p$ is the number of distinct pairs of pulsars. $\bar{r}$ and
$\bar{\zeta}$ indicate the mean values over all pairs of pulsars and
$\sigma_r^2$ and $\sigma_\zeta^2$ are the variances of $r$ and $\zeta$
respectively. For $M$ pulsars, $N_p = M(M-1)/2$.

From the definition of $r(\theta)$ and Eqn. \ref{rho_def}, one can show that the
expected value of $\rho$ is approximately:
\begin{eqnarray}
\rho &\approx& \frac{\sigma_g^2 \sigma_{\zeta}}{\sqrt{\sigma_g^4 \sigma_\zeta^2 + \sigma_{\Delta r}^2}}\\
\sigma_{\Delta r}^2 &=& \frac{1}{N_p} \sum_{i=0}^{Np-1} \langle (r(\theta_i) - \langle r(\theta_i)\rangle )^2\rangle.  \label{sDr}
\end{eqnarray}
For the case where there is no correlation in the data, the statistics
of $\rho$ will be Gaussian with zero mean and variance given by $\sigma_\rho^2 = 1/N_p = 2 / (M^2-M)$.
Hence, the significance of a measured value of $\rho$ may be defined as $S = \rho/\sigma_\rho$.
The probability of measuring a correlation greater than or equal to
$\rho$ when no actual correlation is present is given by
$\mbox{erf}(S/\sqrt{2})/2$.

\section{Estimating the Detection Significance}

In order to estimate the expected detection significance, $S$, one
needs to estimate $\sigma_g$ and $\sigma_{\Delta r}$. It is
assumed that the timing residuals, $R(t,\hat{k})$, are stationary
Gaussian random variables that are sampled at regular intervals
denoted by $\Delta t$. It is also assumed that terms proportional to $t$ and
$t^2$ (i.e.,  the period and period-derivative terms) have been
subtracted from $R(t,\hat{k})$.

The space-time fluctuations induced by a stochastic GW background are
described by a quantity known as the characteristic strain spectrum
denoted by $h_c$ \citep[e.g.][]{mag00}. Models of the GW background propose a
power-law dependence between $h_c$ and the GW frequency, $f$: $h_c(f) = A f^{\alpha}$ \citep{jb03,wl03,mag00,ein+04}.
Using this form of the characteristic strain spectrum, the power
spectrum of the induced residuals is given by $P_R(f) = \langle |\tilde{R}(f)|^2\rangle  = \frac{A^2}{4 \pi^2} f^{2\alpha -3}$,
where $\tilde{R}(f)$ is the Fourier transform of $R(t)$. 
Given $P_R(f)$, the total RMS fluctuation induced by the
stochastic GW background is given by
\begin{eqnarray}
\sigma_g^2 &=&  \int_{f_l}^{f_h} P_R(f) df \\
           &=&  \frac{A^2}{2 \pi^2 (2-2\alpha)} \left(f_l^{2\alpha -2} - f_h^{2\alpha -2}\right) \label{sigmag2}
\end{eqnarray}
where $f_l$ is the lowest detectable frequency given by $1/T$ and
$f_h$ is the highest detectable frequency typically given by
$1/2\Delta t$. $T$ is the total time span of the data set. Since $\alpha <0$ for backgrounds of interest \citep{mag00}, the term containing $f_h$ is negligible.
  
Estimating $\sigma_{\Delta r}$ is slightly more complicated.  To take
into account the effects of subtracting linear and quadratic terms
from the residuals, a semi-analytic approach was adopted. As outlined
below, an estimate for $\sigma_{\Delta r}$ is made analytically but
with one free parameter $\beta$. For a given value of $\beta$, $S$ is
calculated as a function of $A$ for a given set of pulsars and timing
parameters. $S(A)$ is compared to Monte-Carlo simulations in order to
determine the correct value of $\beta$. This showed that the value of
$\beta$ is insensitive to the values $\alpha$, N, M, $\sigma_g$ and
the RMS residual noise level.

Using Equation~\ref{r_stat} together with the assumption that
$R(t,\hat{k})$ is a Gaussian random variable, one can show that
\begin{eqnarray}
\sigma_{\Delta r}^2 &\approx& \overline{\frac{1}{N^2}\sum_{i=0}^{N-1}\sum_{j=0}^{N-1} c_{ij}(\hat{k}_1) c_{ij}(\hat{k}_2)}, \label{eqnsd}
\end{eqnarray}
where $c_{ij}(\hat{k}) =\langle R(t + i \Delta t,\hat{k})R(t+ j \Delta t,\hat{k})\rangle$. The bar above Equation~\ref{eqnsd} represents an average over all
pairs of pulsars.  As the autocorrelation function and the power
spectrum are Fourier transforms of one another, one can estimate
$\sigma_{\Delta r}^2$ from the expected power spectrum of the
residuals. The statistics of the residuals are assumed to be
stationary so that $c_{ij}(\hat{k})$ depends only on $i - j$. The
expected discrete power spectrum of $R(t,\hat{k})$, which includes both a GW component and a white noise component, is given by
\begin{equation}
P_d(i,\hat{k}) = \left\{ \begin{array}{lc} 
          P_g(i) + \frac{2 \sigma_n(\hat{k})^2}{N} \hspace{20pt} & \mbox{for $i > 0$} \\
	  0 & \mbox{for $i = 0$} 
	  \end{array} 
\right.. \label{p_d}
\end{equation}
$P_g(i)$ is the discrete power spectrum of the GW-induced timing
residuals, $i$ is the discrete frequency bin number corresponding to
frequency $i/T$, $\sigma_n(\hat{k})$ is the RMS value of the
residual fluctuations caused by all non-GW sources for the pulsar in
the $\hat{k}$ direction. It is assumed that all noise sources have a
flat spectrum. This assumption is consistent with most observations of
MSPs. $P_g(i)$ is given by
\begin{equation}
P_g(i) =  \frac{A^2 T^{2-2\alpha}}{(2 \pi)^2 (2-2\alpha)}m(i)
\end{equation}\label{pg}
where
$$\begin{array}{ll} 
  m=0 & {\rm for~} i = 0\\
  m= \beta^{2\alpha -2} - (1.5)^{2\alpha-2} & {\rm for~} i = 1  \\
  m=(i-0.5)^{2\alpha-2} - (i+0.5)^{2\alpha-2} & {\rm for~} i > 1.
\end{array}$$
Effectively, $\beta$ is the lowest frequency used to calculate the
correlation function $c_{ij}$. Monte-Carlo simulations show that
$\beta = 0.97$.

For the case where all pulsars have the same noise level, the detection significance becomes 
\begin{eqnarray}
S &=& \sqrt{\frac{M(M-1)/2}{1 + \frac{\chi + 2(\sigma_n/\sigma_g)^2 + (\sigma_n/\sigma_g)^4}{N \sigma_\zeta^2}}}\label{sig_nolambda}
\end{eqnarray}
where $\chi = \frac{1}{\sigma_g^4 N}\sum_{i=0}^{N-1} \sum_{j=0}^{N-1}c_{g_{ij}}^2$, and $c_{g_{ij}}$ is the correlation function for the GW-induced
component of the timing residuals. $\chi$ is a measure of the
``whiteness'' of the residuals. 

\begin{figure}

\plotone{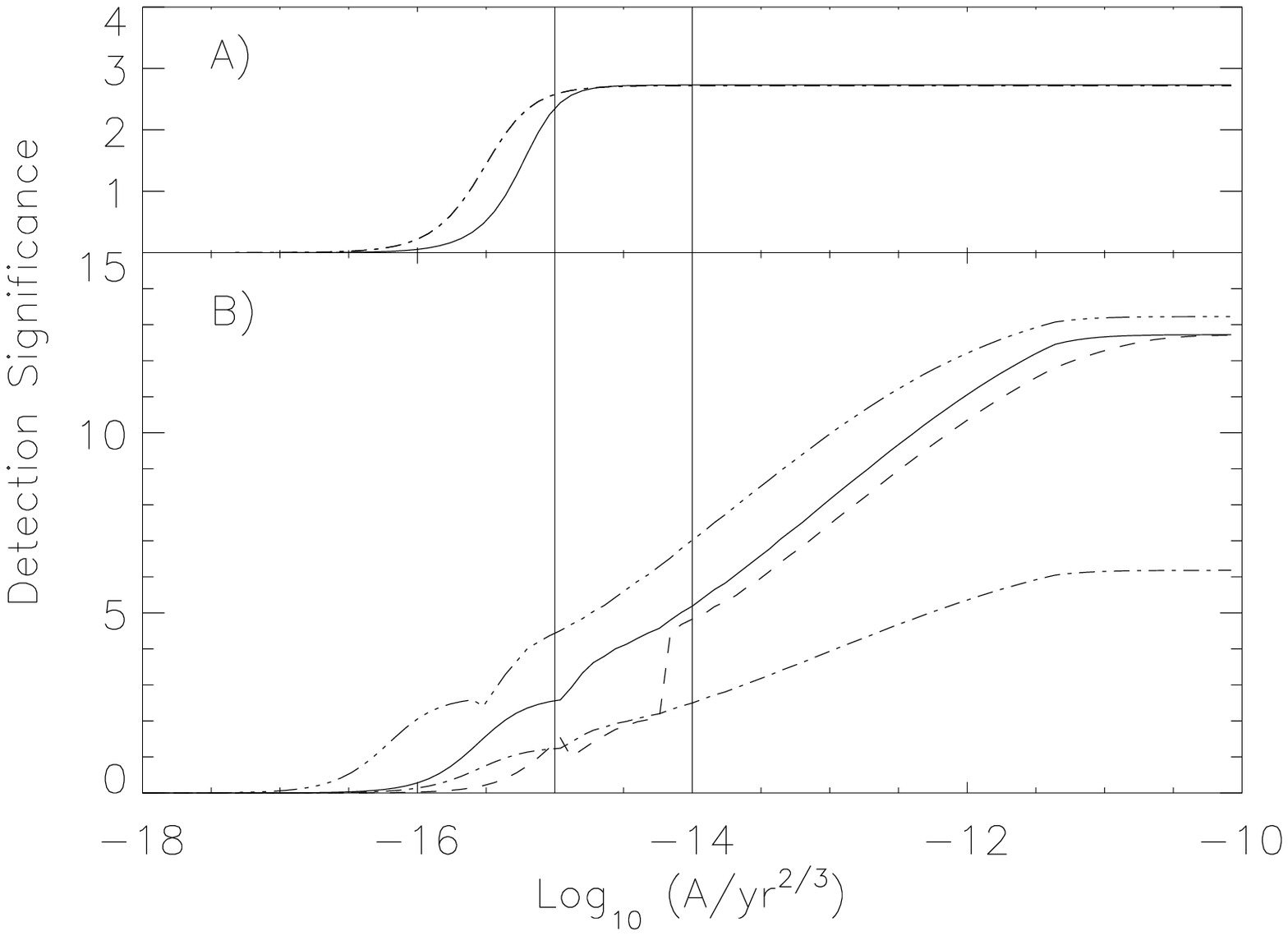}
\caption{\label{fig1} The detection significance, S, versus the
  logarithm of the amplitude $A$ of the characteristic strain
  amplitude $h_c(f)$. The strain spectral index $\alpha = -2/3$,
  corresponding to an astrophysical background of GWs generated by
  super-massive binary black holes. The vertical lines bound the
  values of $A$ expected by models of the GW background
  \citep{jb03,wl03,ein+04}. In panel A), the curves were calculated
  with 20 pulsars each with RMS residual noise fluctuations of 100
  ns. The solid line corresponds to the simple correlation
  technique. The dashed line includes the effect of low-pass
  filtering. Panel B) shows the effects of the whitening
  technique. The solid line was calculated with the same parameters as
  in A). The remaining curves were generated using different noise
  levels and number of pulsars. See text for further details.}

\end{figure}

The solid curve in Figure~\ref{fig1} panel A) plots the detection
significance versus power-law amplitude for $\alpha = -2/3$, the
expected value for a background generated by an ensemble of
super-massive black hole binaries \citep{jb03}. This spectral index
together with the removal of the linear and quadratic terms from $R$
effectively makes $\chi = 0.6 N$. The parameters are set as follows:
$N=250$, $M=20$, $\sigma_n=100$ ns and $T = 5$ years. These values are
the target values for the PPTA \citep{hob04}. Note that the
significance saturates for high values of $A$. This effect can easily
be seen in Equation~\ref{sig_nolambda} since all terms of the form
$\sigma_n/\sigma_g$ go to zero as $\sigma_g$ gets very large. This
saturation is due to the ``self-noise'' associated with the stochastic
nature of the background and its asymptotic value is independent of
$\sigma_n$. The roll-off at low values of $A$ occurs at $\sigma_g =
\sigma_n$.


Since the power spectrum of the GW-induced timing residuals will be
dominated by low frequencies, one can apply a low-pass filter to each
of the residual time series before correlating. This is similar to
fitting a low-order polynomial to the data and then correlating the
resulting fits. To estimate the significance for this technique, one
evaluates $\sigma_g^2$ and $\sigma_{\Delta r}^2$ using
Equations~\ref{sigmag2} and \ref{p_d} but with a high frequency
cut-off $f_{hc}$. For purposes of this discussion, $f_{hc}$ was set to
$4/T$. The dashed line in Figure~\ref{fig1} panel A) shows the effect
of using a low-pass filter on the residuals. All the other parameters
are the same as those used to generate the solid line. Low-pass
filtering effectively reduces $\sigma_n$ while keeping $\sigma_g$
relatively unchanged. It also has the effect of increasing $\chi/N$
when $P_g$ is a red power-law spectrum. Hence, low-pass filtering will
not increase the maximum attainable significance, but it will lower
the value of $\sigma_g$ where the roll-off starts to occur.

We next try to increase the maximum achievable significance. This
method involves both low-pass filtering and a technique called
``whitening''. When correlating two time series that each have a steep
power-law spectrum, an optimal signal-noise ratio is obtained if filters
are applied to give each time series a flat spectrum before
correlation. This will act to reduce $\chi$ in
Equation~\ref{sig_nolambda}. In practice, starting from the lowest
non-zero frequency bin, we give each Fourier component with
significant power equal amplitude and set higher components to zero.
In this way, we are correlating only that part of the signal which has
a high signal-to-noise ratio and adjusting the power spectrum to
optimize the measurement of the correlation function.

$P_d$ and $\sigma_g$ need to be calculated in order to estimate
$S$ using the whitening method. After whitening $P_d(i,\hat{k}) = 2
\sigma_d(\hat{k})^2 / N$, where $\sigma_d(\hat{k})$ is the
RMS of the residual data from the $\hat{k}$th
pulsar. The whitening also affects $\sigma_g$. In
the general case where the pulsars have different noise levels,
$\sigma_g$ will depend on the pulsar. The expression for $\rho$ then
becomes:
\begin{equation}
 \rho \approx \frac{\left(\overline{\sigma_g^2\zeta^2} - \overline{\sigma_g^2\zeta} \mbox{ }  \overline{\zeta} \right)/\sigma_\zeta}{\sqrt{\left(\overline{ \sigma_g^4 \zeta^2} - \left(\overline{\sigma_g^2\zeta}\right)^2\right) + \sigma_{\Delta r}^2}} \label{rho}\\
\end{equation}
with $\sigma_g(\theta)^2$ given by
\begin{eqnarray}
\sigma_{g}(\theta)^2 &=& \frac{2}{N} \sigma_d(\hat{k}_1)\sigma_d(\hat{k}_2) \sqrt{\left(\sum_{i=0}^{N_{max}}P_g(i)/P_d(i,\hat{k}_1)\right)\left(\sum_{i=0}^{N_{max}}P_g(i)/P_d(i,\hat{k}_1)\right)}
\end{eqnarray}
where $N_{max}$ is the largest
frequency bin used based on the criterion discussed above. The
solid line in Figure~\ref{fig1} panel B) plots the significance using the
whitening versus $A$. The same parameters were used for this case as
in the previous cases.

The above discussion assumes that the noise levels were the same for
all pulsars. Next, the case where the pulsars have different noise
levels will be considered. All curves in Figure~\ref{fig1} panel B)
were generated using the whitening technique. Unless specified, 250
observations were taken on each pulsar over 5 years. The dashed line
corresponds to 20 pulsars, 10 with $\sigma_n=100$~ns and 10 with
$\sigma_n=500$~ns. The dashed-dot line has 10 pulsars each with
$\sigma_n=100$~ns and 500 observations. The dashed-triple-dot line has
20 pulsars with $\sigma_n=100$~ns and 500 observations over ten years.

When given a choice between observing a large sample of pulsars with
different noise levels and observing only those pulsars with the
lowest noise levels but for a longer time, the above curves
demonstrate that one should actually observe the larger sample of
pulsars. This is not a general statement, but rather it depends on the
level of the GW background and the noise level. However, the levels
chosen above are relevant to the PPTA
\citep{jb03,wl03,hob04}. Note that for large $M$, the
significance scales as $M$. Hence, doubling the number of pulsars will
double the expected significance.

\section{Summary}

The main goal of this work is to determine the effectiveness of an
array of pulsars, such as the PPTA, for detecting a stochastic
background of GWs. Using a simple correlation technique, the detection
significance was calculated given the number of pulsars, the location
of each pulsar, the TOA precision, the number of observations, the
total time span of the data, and the amplitude and power-law index of
the GW background. For the case where all pulsars have the same
white-noise spectrum, Equation~\ref{sig_nolambda}, may be used to
calculate the detection significance. For the case of the PPTA, it was
found that the maximum achievable significance will be about 3 for a
background with spectral index $\alpha = -2/3$ and $A \sim 10^{-15}$
which is the expected level of the GW background from an
ensemble of super-massive binary black holes in galaxies
\citep{jb03,wl03,ein+04}. Note that lowering the RMS noise level will only
decrease the minimum detectable value of $A$ and not increase the
maximum attainable significance.

Low-pass filtering the timing residuals, or equivalently, fitting
low-order polynomials (i.e. cubic terms) to the residuals and
correlating the coefficients, does not increase the maximum attainable
significance. The significance level is increased by pre-whitening of
the timing residuals. Using whitening, it is estimated that the PPTA
could obtain a detection significance greater then 4 for $A \geq 3
\times 10^{-15} \mbox{yr}^{-2/3}$ provided that efficient whitening
filters can be designed and implemented. This is an area of further
study and will be addressed in a later paper. With the same
qualifiers, increasing the total time span of the PPTA to 10 years
would yield a significance greater then 4 for $A \geq 10^{-15}
\mbox{yr}^{-2/3}$. Since the significance scales as the number of
pulsars, doubling that number will double the expected
significance. Hence, using the simple correlation technique described
here without any pre-filtering, a stochastic background with $A \geq
10^{-15} \mbox{yr}^{-2/3}$ will be detectable at a significance of about 5.5
using 40 pulsars observed 250 times over 5 years and each having
100~ns timing precision.

Part of this research was carried out at the Jet Propulsion
Laboratory, California Institute of Technology, under a contract with
the National Aeronautics and Space Administration and funded through
the internal Research and Technology Development program. The authors
wish to thank Russell Edwards for useful discussions.

\end{document}